\documentclass[sigconf]{acmart}

\usepackage{amsfonts}

\usepackage{bm}
\usepackage{array}
\usepackage{color}
\usepackage[normalem]{ulem}
\usepackage{colortbl}
\usepackage{textcomp}
\usepackage{stfloats}
\usepackage{hyperref}
\usepackage{verbatim}
\usepackage{graphicx}
\usepackage{subfig}
\usepackage{amsmath}

\usepackage{amssymb}

\usepackage{latexsym}
\usepackage{xspace}
\usepackage[T1]{fontenc}
\usepackage{bbding}
\usepackage{pifont}
\usepackage[utf8]{inputenc}

\usepackage{microtype}

\usepackage{paralist}

\newcommand{\paratitle}[1]{\vspace{1.4ex}\noindent \textbf{#1}}

\newcommand{\logoWiki}{\raisebox{-3pt}{\includegraphics[width=1.3em]{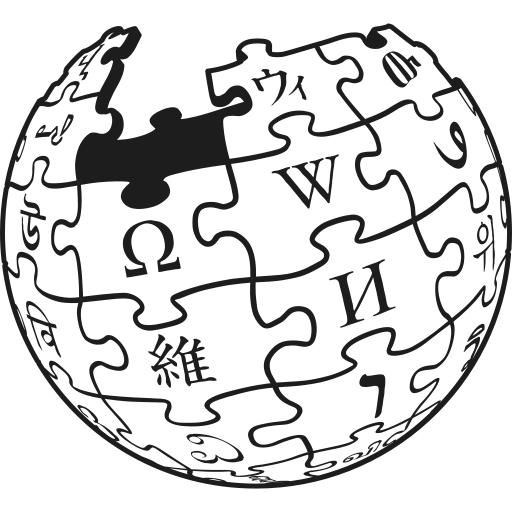}}\xspace}

\newcommand{\logoMedicine}{\raisebox{-3pt}{\includegraphics[width=1.3em]{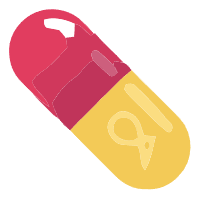}}\xspace}

\newcommand{\logoGene}{\raisebox{-3pt}{\includegraphics[width=1.3em]{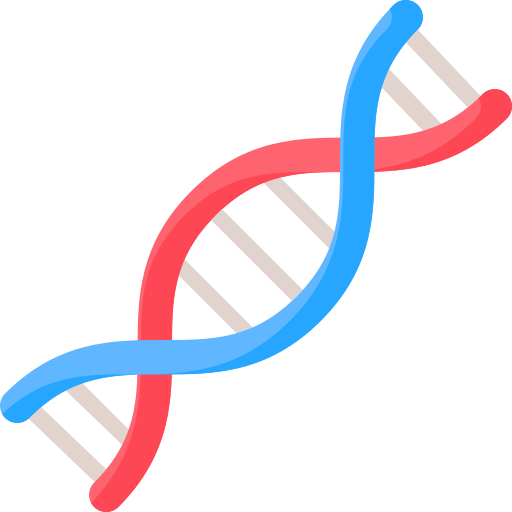}}\xspace}

\newcommand{\logoNews}{\raisebox{-3pt}{\includegraphics[width=1.3em]{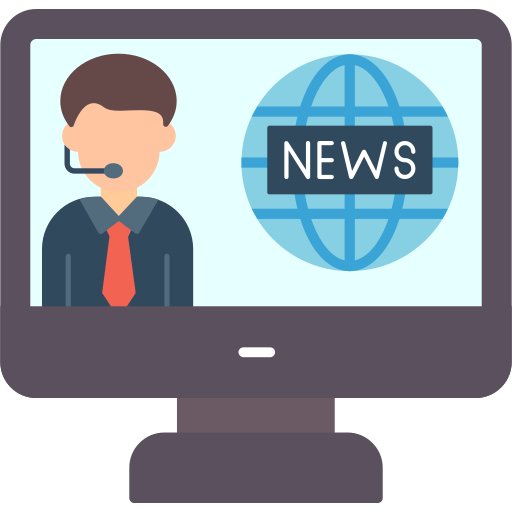}}\xspace}

\newcommand{\logoMovie}{\raisebox{-3pt}{\includegraphics[width=1.4em]{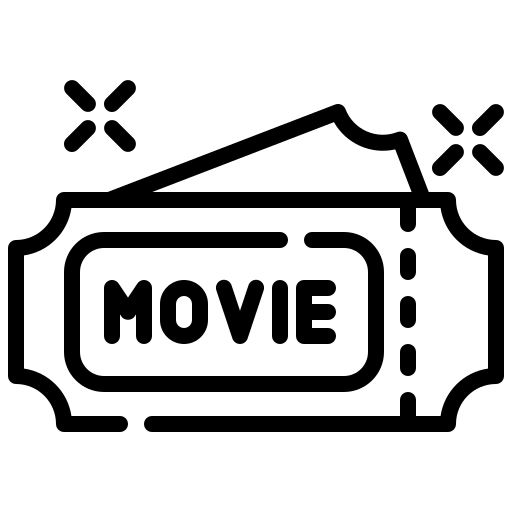}}\xspace}

\newcommand{\logoCyber}{\raisebox{-2pt}{\includegraphics[width=1.2em]{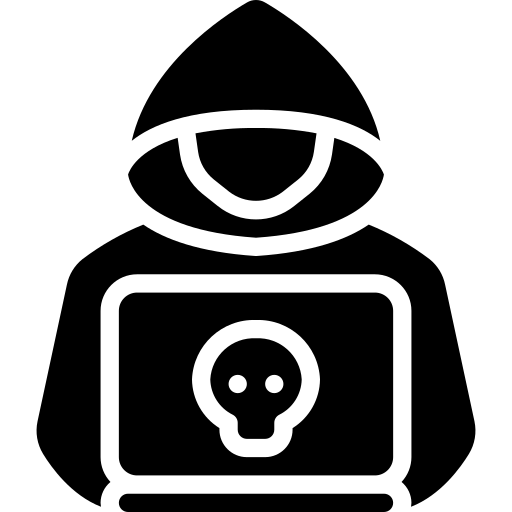}}\xspace}

\definecolor{lightblue}{rgb}{0.93, 0.95, 1.0}

\allowdisplaybreaks 

\usepackage{multirow} 
\usepackage{booktabs} 
\usepackage{arydshln} 


\AtBeginDocument{%
  }

\setcopyright{acmlicensed}

\copyrightyear{2024}
\acmYear{2024}
\setcopyright{acmlicensed}\acmConference[MM '24]{Proceedings of the 32nd ACM International Conference on Multimedia}{October 28-November 1, 2024}{Melbourne, VIC, Australia}
\acmBooktitle{Proceedings of the 32nd ACM International Conference on Multimedia (MM '24), October 28-November 1, 2024, Melbourne, VIC, Australia}
\acmDOI{10.1145/3664647.3680669}
\acmISBN{979-8-4007-0686-8/24/10}

\begin{document}

\title[Speech Event Extraction]{SpeechEE: A Novel Benchmark for Speech Event Extraction}


\author{Bin Wang}
\affiliation{%
  \institution{Harbin Institute of Technology (Shenzhen)}
  \city{Shenzhen}
  \country{China}
  }
\email{23s051047@stu.hit.edu.cn}

\author{Meishan Zhang}
\affiliation{%
  \institution{Harbin Institute of Technology (Shenzhen)}
  \city{Shenzhen}
  \country{China}
  }
\email{zhangmeishan@hit.edu.cn}

\author{Hao Fei}
\authornote{Hao Fei is the corresponding author.}
\affiliation{%
  \institution{National University of Singapore}
  \city{Singapore}
  \country{Singapore}
  }
\email{haofei37@nus.edu.sg}

\author{Yu Zhao}
\affiliation{%
  \institution{Tianjin University}
  \city{Tianjin}
  \country{China}
  }
\email{zhaoyucs@tju.edu.cn}

\author{Bobo Li}
\affiliation{%
    \institution{Wuhan University}
  \city{Wuhan}
  \country{China}
}
\email{boboli@whu.edu.cn}

\author{Shengqiong Wu}
\affiliation{%
  \institution{National University of Singapore}
  \city{Singapore}
  \country{Singapore}
  }
\email{swu@u.nus.edu}

\author{Wei Ji}
\affiliation{%
  \institution{National University of Singapore}
  \city{Singapore}
  \country{Singapore}
  }
\email{jiwei@nus.edu.sg}

\author{Min Zhang}
\affiliation{%
  \institution{Harbin Institute of Technology (Shenzhen)}
  \city{Shenzhen}
  \country{China}
  }
\email{zhangmin2021@hit.edu.cn}

\renewcommand{\shortauthors}{Bin Wang et al.}

\begin{abstract}
Event extraction (EE) is a critical direction in the field of information extraction, laying an important foundation for the construction of structured knowledge bases.
EE from text has received ample research and attention for years, yet there can be numerous real-world applications that require direct information acquisition from speech signals, online meeting minutes, interview summaries, press releases, etc.
While EE from speech has remained under-explored, this paper fills the gap by pioneering a \textbf{SpeechEE}, defined as detecting the event predicates and arguments from a given audio speech.
To benchmark the SpeechEE task, we first construct a large-scale high-quality dataset.
Based on textual EE datasets under the sentence, document, and dialogue scenarios, we convert texts into speeches through both manual real-person narration and automatic synthesis, empowering the data with diverse scenarios, languages, domains, ambiences, and speaker styles.
Further, to effectively address the key challenges in the task, we tailor an E2E SpeechEE system based on the encoder-decoder architecture, where a novel Shrinking Unit module and a retrieval-aided decoding mechanism are devised.
Extensive experimental results on all SpeechEE subsets demonstrate the efficacy of the proposed model, offering a strong baseline for the task. 
At last, being the first work on this topic, we shed light on key directions for future research. 
Our codes and the benchmark datasets are open at \url{https://SpeechEE.github.io/}.
\end{abstract}


\ccsdesc[500]{Information systems~Multimedia information systems}

\keywords{Information Extraction, Event Extraction, Speech Modeling, Spoken Language Understanding}

\maketitle

\section{Introduction}

Event extraction \cite{EE_intro1,EE_intro2} is a critical task within the information extraction community \cite{li2022unified,fei2022lasuie}, aimed at automatically identifying structured information from various data sources. 
It seeks to delineate the semantic structure encapsulating the essence of `\emph{who or what does what to whom, when, where, and why}' \cite{EE}. 
Initially, research in EE predominantly focuses on textual data \cite{textee}; and over time, it became evident that events could be conveyed through a myriad of modalities and information sources. 
Subsequently, the scope of EE has broadened to include more diverse modalities and information sources, leading to significant strides in extracting events from images \cite{m2e2} and even videos \cite{video}.
Despite these advances, extracting events from speech or audio signals remains a largely under-explored topic.
We argue that EE in speech also holds immense research significance and practical value, given its applicability in a variety of real-life scenarios, including meetings, lectures, interviews, and news reports, especially in scenarios where transcription is not available.

\begin{figure}[!t]
\centering
\includegraphics[width=0.78\columnwidth]{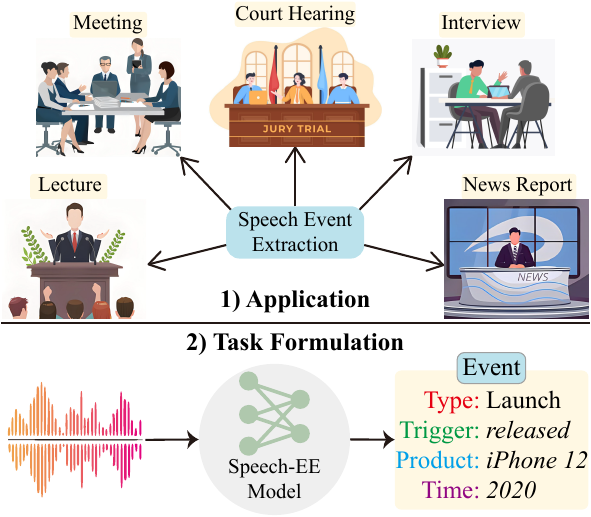}
\vspace{-2mm}
\caption{
An illustration of the speech event extraction's broad applications and task formulation.
}
\label{fig:intro}
\vspace{-5mm}
\end{figure}

In response to this gap, in this paper, we introduce a novel task: Speech Event Extraction (namely, \emph{\bf SpeechEE}). 
SpeechEE is designed to process audio inputs and output structured event records, identifying event triggers, categorizing event types, recognizing arguments and classifying their roles.
To benchmark this task, we develop a large-scale and high-quality comprehensive dataset.
On the one hand, based on 8 common textual EE datasets under the \emph{sentence}, \emph{document} and \emph{dialogue} upon ACE EE annotation format \cite{EE}, we meticulously convert texts into diverse and authentic speeches through manual real-person narration, where we simulate environments with both quiet and noisy backgrounds.
To further enlarge the data quantity, we automatically synthesize the SpeechEE data via state-of-the-art (SoTA) text-to-speech (TTS) systems, progressively extending the amount while preserving all the characteristics.
Strict human cross-inspection is conducted to ensure the high quality of the whole speech data. 
The final SpeechEE dataset comprises 8 subsets, featuring diverse
1) \textbf{scenarios} (sentences, documents, dialogues), 
2) \textbf{languages} (English and Chinese), 
3) \textbf{domains} (news, cybersecurity, movies, etc.), 
4) \textbf{ambiences} (noisy and quiet) and 
5) \textbf{speaker styles}. 
Our experimental analyses reveal that the SpeechEE dataset poses greater challenges compared to traditional textual EE, underscoring the complexity and uniqueness of speech as a medium for EE.

Modeling SpeechEE presents indispensable challenges. 
The most straightforward approach is first converting speech to text using Automated Speech Recognition (ASR) tools \cite{w2v2}, followed by applying existing textual EE techniques \cite{ONEEE}. 
However, this pipeline approach suffers from significant error propagation issues. 
More crucially, it fails to address several key bottlenecks inherent to SpeechEE.
\textbf{First}, speech inherently flows without clear word boundaries, presenting a challenge in accurately identifying the precise audio segments associated with event triggers and arguments.
\textbf{Second},
in real scenarios, speech might encompass background noise, hindering the effective extraction of event-relevant semantic features (e.g., triggers and arguments) directly from the characteristics of the speech itself.
\textbf{Third}, the length of audio signals can be significantly greater than their text counterparts, often by orders of magnitude, adding complexity to the modeling of speech-to-event extraction.
\textbf{Finally}, the presence of homophones and near-homophones in speech can lead to inaccuracies in entity recognition.
For instance, words like `peace' vs. `piece' or `male' vs. `mail' can lead to inaccuracies, particularly with uncommon nouns, such as rare names or locations. 
Capturing these nuances accurately poses a significant challenge for models.

To effectively tackle these challenges, we propose a novel E2E model for SpeechEE task, which has been shown in Fig. \ref{fig:architecture}.
First, our model leverages an encoder-decoder generative framework \cite{Text2Event} to directly produce target event schemes from speech, thereby avoiding the need for piecemeal speech segmentation. 
Then, we employ contrastive learning \cite{CL} at the encoder stage for event representation learning, enhancing the model's ability to discern and disentangle event semantics from speech features. 
Further, a Shrinking Unit module is designed to alleviate the disparities in modal length between speech and text through projection and downsampling techniques.
Finally, our model incorporates a retrieval-aided decoder that leverages an external Entity Dictionary, enabling flexible decision-making during decoding to generate new tokens or retrieve entities directly from the dictionary. 
Our experiments on the SpeechEE dataset demonstrate that our model outperforms pipeline baseline systems consistently. 
Further analysis confirms the substantial contributions of all the proposed components in effectively modeling speech for EE.

In summary, our contributions are threefold: 

\noindent$\bullet$ We for the first time pioneer a novel task, SpeechEE, for extracting events from speech, along with the first large-scale high-quality dataset encompassing multiple scenarios, domains, languages, ambiences, and styles from both synthesis and manual crafting. 

\noindent$\bullet$ We propose an E2E SpeechEE model that addresses key challenges in SpeechEE and offers a strong baseline performance on the benchmark data. 

\noindent$\bullet$ We outline potential future directions to further advance the field of SpeechEE, setting the stage for ongoing research and development in this promising area.

\vspace{-3mm}
\section{Related Works}

EE is one of the kernel subtopics within the field of information extraction \cite{IE,EE_related_work1,EE_intro2,EE_related_work3,EE_related_work4,EE_related_work5}, and has been the subject of extensive research for many years. 
The majority of EE research has focused on textual EE \cite{Text2Event,ONEEE} due to the abundance of text information available online. 
As one of the most popular tasks in natural language processing, EE has evolved over decades and garnered significant research attention. 
A variety of EE methods have been proposed, such as classification approaches \cite{EE_intro2}, sequence labeling techniques \cite{seqlabel}, question-answering formats \cite{QA}, and generation methods \cite{GA}, etc. 
Given its critical applications across numerous scenarios, EE has been integrated into various modalities \cite{m2e2,video}, such as image and video EE \cite{wu2023information,zhang2024recognizing}.
On the other hand, while research on Named Entity Recognition (NER) \cite{speechNER} and Relation Extraction (RE) \cite{ren2022simple,Re_related,SpeechRE} under text and speech have emerged within the IE community, research specifically addressing EE under speech remains conspicuously absent. 
Thus, this work endeavors to fill this gap by establishing a comprehensive SpeechEE benchmark dataset. 
We develop a large-scale dataset through both manual and automated methods, broadly encompassing multiple scenarios, languages, domains, ambiences, and speaker styles.

This work also relates to the speech modeling research \cite{speech_modeling1,speech_modeling2,speech_modeling3}. 
Audio signals represent a significant modality in the world, especially within human society where speech is a primary mode of communication, which is one of the key research tracks within the multimodal learning communities \cite{fei2022matching,li2022fine,fei2023scene,wu2023cross2stra,li2023revisiting}, e.g., image modeling \cite{ji2021improving,ji2022knowing,wu2023imagine,li2023variational}, video modeling \cite{zhao2023constructing,meng2024mmlscu,fei2024dysen,fei2024video}.
Consequently, the speech community has focused on various tasks, directions, and research scenarios, including ASR \cite{w2v2}, TTS \cite{TTS}, and Spoken Language Understanding (SLU) \cite{SLU}, among others.
To enable machines to comprehend the semantic information in speech, particularly to extract linguistic information such as event structures, accurate understanding of speech is required. 
The conventional approach \cite{SpeechRE} involves first using ASR technology to transcribe speech into text, followed by the application of established NLP techniques for textual analysis. 
However, this pipeline approach inevitably introduces distortions in information extraction from the given speech due to potential errors in ASR, thereby incorporating noise \cite{Error_propagation}. 
In this paper, we construct an E2E SpeechEE model that addresses a series of unique challenges associated with SpeechEE.

\vspace{-2mm}
\section{Task Definition of SpeechEE}

We mainly follow the ACE scheme \cite{EE} for the event definition.
We formulate the SpeechEE task as: 
given a speech audio consisting of a sequence of acoustic frames $\boldsymbol{S} = (\boldsymbol{f_1},\boldsymbol{f_2},\cdots,\boldsymbol{f_U})$, the pre-defined event type set $E$ and argument role set $R$, we aim to extract all possible structured event records consisting of four parts: 
1) event type $\varepsilon \in E$, 
2) event trigger, 
3) event argument role $r \in R$ 
and 4) event argument.

\vspace{-1mm}
\section{Benchmark Data Construction}
\label{sec:dataset}

\subsection{Construction Approach}

\paratitle{Data Source.}
We consider constructing our SpeechEE data based on existing textual EE benchmark datasets, since textual EE datasets are well-defined and well-established, with readily available and accurate event annotations. 
Specifically, we utilize datasets from three scenarios:
1) sentence-level data, including ACE05-EN\textsuperscript{+} \cite{metrics3}, ACE05-ZH \cite{ACErawdata}, PHEE \cite{PHEE}, CASIE \cite{CASIE}, and GENIA \cite{GENIA};
2) document-level data, RAMS \cite{RAMS} and WikiEvents \cite{Wiki};
and 3) dialogue-level data, Duconv subset in CSRL \cite{CSRLdata}.
Each dataset strictly follows the ACE EE schema and is annotated with corresponding EE records.

\vspace{-1mm}
\paratitle{Manual Speech Narration.}
Based on the above textual data, we then obtain the corresponding speech signals through manual reading. 
Note that each dataset also maintains its original train/dev/test split, which we do not alter.
For each dataset's respective languages, we employ 10 native speakers of different genders and ages to ensure speech diversification in terms of tone and timbre. 
Each speaker is instructed to read the original text data in both quiet and noisy background settings, such as in a cafeteria, meeting room, street, and classroom, to cover as many real-life scenarios as possible. 
To ensure the quality of the acquired speech, we conduct manual cross-inspection. 
Specifically, 2 individuals simultaneously listen to the same speech recording and score it from 1-10 (low to high quality) based on the audio's accuracy in reflecting the original text. 
Finally, we calculate the Cohesion Kappa score across the 2 auditors, retaining only instances where the score exceeded \textbf{0.85}, thereby ensuring high consistency in data quality.

\paratitle{Automatic Synthesis.}
Due to the huge workload and cost of manual speech recording, as well as the fact that some scenarios has hard real environment reproduction, we can record only a portion of speech for each original text dataset. 
To substantially expand our SpeechEE dataset, we now consider using automatic synthesis to continue building speech. 
Note that we only enrich the train set of SpeechEE data.
Our basic idea is using TTS tools to convert text to speech automatically, during which we strictly control the synthesized voice's timbre and ambient sounds. 
We mainly use two high-performance open-source TTS tools: Bark\footnote{\url{https://github.com/suno-ai/bark}} and Edge-TTS\footnote{\url{https://github.com/rany2/edge-tts}}. 
Notably, we perform denoising of the original text before auto-recording to filter out instances that are inexpressible in speech or irrelevant to the task, thus ensuring the quality of the resulting speech.
To ensure the quality of the synthesized data, we consider different evaluation methods from those used for manual construction. 
Specifically, we evaluate from both objective and subjective perspectives: the former assesses the text accuracy of synthesis speech, and the latter evaluates such as the naturalness, timbre, and accuracy of ambient sounds of the speech. 
Objectively, we use an ASR model to evaluate the word error rate of the synthesis speech. 
Subjectively, we employ two native speakers to rate the speech on a 10-scale, retaining only instances where the score exceeds \textbf{0.8}.

\begin{table}
\centering
\caption{
Key characteristics of our SpeechEE dataset. 
}
\vspace{-3mm}
\fontsize{7}{10}\selectfont
\setlength{\tabcolsep}{0.8mm}

\begin{tabular}{llllccc}
\hline

\textbf{Scenario} & \textbf{Source} & \textbf{Language} & \textbf{Domain} & \textbf{Tone} &

\textbf{Event-Type} & \textbf{Arg.-Role} \\

\hline

\multirow{5}{*}{Sentence} & ACE05-EN\textsuperscript{+}  
&English & \logoNews News   &10   &33 &22  \\

&ACE05-ZH & 
Chinese & \logoNews News &  6  &33 &27   \\

&PHEE  
&English & \logoMedicine Medical  & 10 &2 &16  \\

&GENIA 
&English & \logoGene Biology  & 10 &5 &- \\ 

&CASIE
&English & \logoCyber Cyber &10 & 5 &26    \\
\hline

\multirow{2}{*}{Document}
& RAMS	&English & \logoNews  News &7 & 139 &65   \\

& WikiEvents &English & \logoWiki General & 7 & 50 &59   \\

\hline
\multirow{1}{*}{Dialogue} & Duconv 	 &Chinese & \logoMovie Movies  & 6
& 1 & 8   \\
\hline
\end{tabular}

\label{tab:data table}%
\end{table}%

\begin{table}
 \centering
\caption{
Statistics of the SpeechEE dataset. 
In the brackets are the splits of train/develop/test sets.
}
\vspace{-3mm}
\fontsize{7}{9.5}\selectfont
\setlength{\tabcolsep}{3mm}
\begin{tabular}{llcc}
\hline

\textbf{Scenario} & \textbf{Source}  &  \textbf{Human}  &\textbf{Synthesis}  \\

\hline

\multirow{5}{*}{Sentence} & ACE05-EN\textsuperscript{+} \cite{metrics3} & 1,000 (800/100/100)   & 12,867 \\

&ACE05-ZH \cite{ACErawdata} &1,000 (800/100/100) & 6,311\\

&PHEE \cite{PHEE}  & 500 (400/50/50) & 2,898  \\

&GENIA \cite{GENIA} &1,000 (800/100/100)  & 15,023 \\

&CASIE \cite{CASIE} &500 (400/50/50)  & 3,751 \\
\cdashline{2-4}
& Total & 4,000  & 40,850 \\
\hline

\multirow{2}{*}{Document} &RAMS \cite{RAMS}	 & 1,000 (800/100/100) & 7,329\\

&WikiEvents \cite{Wiki}	 & 100 (80/10/10) & 206  \\
\cdashline{2-4}
& Total & 1,100  & 7,535 \\

\hline

\multirow{1}{*}{Dialogue} &Duconv \cite{CSRLdata} 	 & 140 (100/20/20) & 1,200  \\
\hline

\end{tabular}

\label{tab:statistics}%
\end{table}%

\begin{figure*}[!t]
    \centering
    \includegraphics[width=0.92\textwidth]{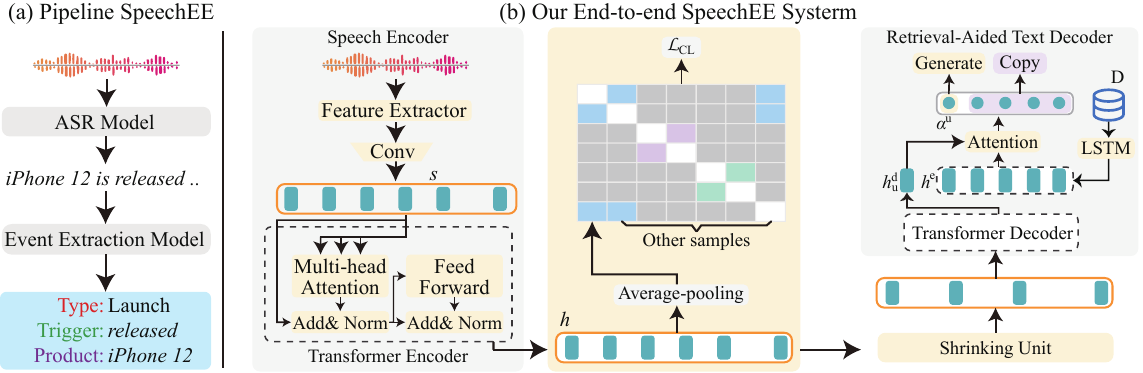}
    \caption{
    The architecture of the pipeline and E2E SpeechEE model. 
    }
    \label{fig:architecture}
   
\end{figure*}

\subsection{Data Insights and Characteristics}

The data characteristics are clearly illustrated in Table \ref{tab:data table}. 
And in Table \ref{tab:statistics} we display the detailed statistics.
Following we summarize the highlights of SpeechEE dataset.

\paratitle{$\bullet$ Multiple Scenarios:} SpeechEE covers three major common scenarios of existing EE: sentence, document and dialogue.

\paratitle{$\bullet$ Diverse Domains:} SpeechEE involves diverse domains, such as news, biomedicine, cybersecurity, movie fashions, etc.

\paratitle{$\bullet$ Multilinguity:} Datasets in SpeechEE cover two languages, English in 6 subsets and Chinese in 2 subsets.

\paratitle{$\bullet$ Various Ambiences:} Speeches either have quiet backgrounds or noisy backgrounds. 
The noisy background setting covers many scenarios that simulate realistic environments of the task.

\paratitle{$\bullet$ Rich Styles:} SpeechEE features a diverse range of human voice styles, e.g., male, female, and child voices, and also different speaker tones and timbres.

\paratitle{$\bullet$ Large Scale and High Quality:} 
We create a large-scale dataset through both manual annotation and automatic synthesis.
The quality has been rigorously controlled through strict cross-validation.

\vspace{-2mm}
\section{SpeechEE Method}

Now we introduce two methods to address SpeechEE, including pipeline SpeechEE and E2E SpeechEE. 
The pipeline SpeechEE is a two-step method that firstly uses ASR system to obtain the transcripts of input speech and then uses textual EE model to extract event records from transcripts.
We then propose an E2E SpeechEE model to extract the event records from speech in one shot.
Two SpeechEE architectures are overviewed in Fig. \ref{fig:architecture}.

\vspace{-2mm}
\subsection{Pipeline Model}

The straightforward way for SpeechEE is to divide it into two subtasks, ASR and textual EE, and cascade strong-performing off-the-shelf models as a two-step pipeline, shown in Fig. \ref{fig:architecture}(a).
Here we provide a feasible implementation.
We first employ the Whisper \cite{whisper} model as the ASR module to convert speech signals to the corresponding transcripts.
Compared with other ASR tools (Wav2Vec 2.0 \cite{w2v2} and HuBERT \cite{HuBERT}), the Whisper model is trained on a considerably labeled audio corpus, and thus can directly learn the mapping from speech to text, with more superior speech recognition performance than other ASR models. 
Additionally, the whisper model incorporates data from multiple languages and domains, which matches the multilingual and diverse domain characteristics of the SpeechEE dataset we construct.
For TextEE module, we adopt Text2Event \cite{Text2Event} which is a generative-based E2E EE method. 
Text2Event features a sequence-to-structure paradigm, which can directly perform textual EE based on the whisper model outputs.
This also helps tackle the lack of fine-grained annotations about the boundary of event trigger and argument mention in SpeechEE.

\subsection{End-to-End Model}

As mentioned previously, the pipeline SpeechEE method faces significant error propagation issues.
To address this,
we propose an end-to-end (E2E) approach.  
As illustrated in Fig. \ref{fig:architecture}(b), the overall framework has an encoder-decoder structure, which mainly consists of a speech encoder, a Shrinking Unit module, and a retrieval-aided text decoder.

\paratitle{Speech Encoder.}
We take the speech encoder as in the Whisper model, which is built based on an acoustic feature extractor and a normal transformer encoder.
The input speech $S$ is firstly processed by an acoustic feature extractor to get an 80-channel log-magnitude Mel spectrogram clip sequence. 
Then a small stem consisting of two convolution layers with a filter width of 3 and the GELU activation function \cite{hendrycks2016gaussian} is applied to transfer the clip sequence to the inputs of transformer $\boldsymbol{s}$.
Afterward, the transformer encoder encodes the spectrogram representation into hidden states $\boldsymbol{h}=(\boldsymbol{h_{1},h_{2},\cdots,h_{T}})$, where $T$ is the sequence length of hidden states $\boldsymbol{h}$.
We use the last encoder layer's hidden vectors as speech representations.
\setlength\abovedisplayskip{3pt}
\setlength\belowdisplayskip{3pt}
\begin{equation}
\boldsymbol{s} = \operatorname{Conv}(\operatorname{FeatureExtractor}(S)) \,,
\end{equation}
\begin{equation}
\boldsymbol{h} = \operatorname{TransformerEncoder}(\boldsymbol{s}) \,.
\end{equation}

The speech representations can adequately capture the acoustic feature due to the ASR pre-training of Whisper. 
However, these speech representations are not capable of modeling the event-related features for the SpeechEE task.
Thus, we design a contrastive learning strategy to enhance speech representation.
To better learn the event-related semantics, we choose the positive and negative samples based on the event type.
For the same event type, the encoded speech representations will be pulled together while representations for different event types should be pushed away.
For a batch of $N$ samples, the contrastive learning loss $\mathcal{L}_{\text{CL}}$ is defined as follows:
\setlength\abovedisplayskip{3pt}
\setlength\belowdisplayskip{3pt}
\begin{equation}\label{eq:cl}
\mathcal{L}_{\text{CL}} = - \sum_{i=1}^{N} \log \frac{\exp{(\boldsymbol{x^{i}} \cdot \boldsymbol{x^{+}}/\tau)}}{\sum_{\boldsymbol{x}\in {\boldsymbol{K}}} \exp{(\boldsymbol{x^{i}} \cdot \boldsymbol{x}/\tau)}} \,,
\end{equation}
where $\boldsymbol{K}$ is a set composed of all samples in the batch, $\boldsymbol{x^{+}}$ denotes the positive sample, and $\tau$ is the temperature hyper-parameter.
The $x^i$ is the average pooling results for the i-th sample's speech representation $\boldsymbol{h}$.

\paratitle{Shrinking Unit.}
Speech sequences are usually much longer than corresponding text sequences, and there exists even more redundant information for the EE task.
The discrepancy in sequence length of different modalities adds great difficulty to the modeling of SpeechEE task.
To combat this, we here propose a novel Shrinking Unit module, which is added between the speech encoder and text decoder in our E2E architecture, to mitigate the difference in the sequence length by projection and downsampling. 
Technically, $n$ one-dimensional convolutional layers downsample the encoder output $\boldsymbol{h}$ using a stride of $m$, which can shorten the sequence by a factor of $m^{n}$.
\setlength\abovedisplayskip{3pt}
\setlength\belowdisplayskip{3pt}
\begin{equation}
\boldsymbol{h_{s}} = \operatorname{Shrinking Unit}(\boldsymbol{h}) \,.
\end{equation}

\paratitle{Retrieval-Aided Text Decoder.}
After the hidden state vectors are filtered by the Shrinking Unit, a pre-trained text decoder predicts the output event structure token-by-token with the sequential input tokens’ hidden vectors. 
At the step $k$ of generation, the text decoder predicts the $k$-th token $y_{k}$
and decoder state $\boldsymbol{h^{d}_{k}}$ as follows:
\begin{equation}
    \boldsymbol{h^{d}_{k}} , y_{k}= \operatorname{Decoder}(\boldsymbol{h},\boldsymbol{h^{d}_{1}},\cdots,\boldsymbol{h^{d}_{k-1}},y_{k-1}) \,.
\end{equation}

Yet due to the phenomenon of homophones and near-sound words in speech, it is hard for the decoder to precisely extract some entity mentions, especially for rarely-seen words during the training.
Inspired by the Contextual ASR \cite{contextualASR}, we propose to incorporate a retrieval mechanism and leverage an external Entity Dictionary to flexibly decide whether to generate a new token or retrieve an existing entity from the Entity Dictionary.
The retrieval mechanism can help to constrain the difficult entities with a closed set and avoid generating incorrect ones with flawed homophones and near-sound words.

Technically, the Entity Dictionary is constructed with entities that appear only once in the training set, covering the rarely-seen words that are hard to recognize.
We assume that we have no other prior knowledge about the test data. Therefore, the train set from the same origin can be used as the external knowledge properly.
We denote the Entity Dictionary as $D=\{e_{0},e_{1},e_{2},\cdots,e_{l}\}$, where $e_{0}$ is added to note the no-entity option. The Entity Dictionary is firstly encoded by an LSTM module to get the last state as the fixed dimensional entity representation $h^{e}=\{\boldsymbol{h^{e}_{0}},\boldsymbol{h^{e}_{1}},\cdots,\boldsymbol{h^{e}_{l}} \}$.
Then the attention score for entity $e_{j}$ is computed where query denotes the last layer of decoder state $\boldsymbol{h^{d}}$ and the key denotes the entity representation $\boldsymbol{h^{e}}$. 
\setlength\abovedisplayskip{3pt}
\setlength\belowdisplayskip{3pt}
\begin{equation}
    \alpha_{j}^{u} = \frac{(\boldsymbol{W_{q}h_{u}^{d}}) \cdot (\boldsymbol{W_{k}h_{j}^{e}})}{\sqrt{d_{att}}} \,,
\end{equation}
where $d_{att}$ denotes the dimension of $\boldsymbol{h^{e}_{j}}$, $u$ denotes the decoding step, $\boldsymbol{W_{q}}$ and $\boldsymbol{W_{k}}$ are two learned linear transformation parameters of query and key respectively.
After softmax, we obtain the retrieved probability $ p_{j}^{u}$ of the entity $e_{j}$. It is used to flexibly decide to retrieve which existing entity in the dictionary or generate the output by decoder. Then we compute the loss by using the golden entity label and the retrieved probability.
\setlength\abovedisplayskip{3pt}
\setlength\belowdisplayskip{3pt}
\begin{equation}
    \mathcal{L}_{\text{ED}} = - \sum_{u} log p_{g}^{u} \,,
\end{equation}
where $g$ denotes the golden entity in time step $u$.
The final loss is composed of $\mathcal{L}_{\text{ED}}$ and the contrastive loss $\mathcal{L}_{\text{CL}}$ in Equation \ref{eq:cl}.

\section{Experiments}

\subsection{Settings}

We carry out experiments on our SpeechEE datasets.
For the pipeline baseline, we use whisper-large-v2 as the ASR module and choose Text2Event with two different language models, i.e., Bart-large and T5-large as the TextEE module.
For our E2E method, for a fair comparison, we also adopt the encoder of whisper-large-v2 as the speech encoder and use the decoder of pre-trained language models (Bart-large and T5-large) as our text decoder.
In particular, we change Bart-large and T5-large to mBart-large-50 and mT5-large for Chinese subsets, ACE05-ZH and Duconv.

For efficient training, we freeze the acoustic feature extraction module of whisper and train the self-attention encoder, Shrinking Unit module, cross-attention between encoder and decoder, and Entity Dictionary attention. 
We train all SpeechEE models and optimize the models using AdamW \cite{Adamw}.
We conduct all the experiments on NVIDIA A100 80GB. 
All of our models are evaluated on the best-performing checkpoint on the validation set.
After the model inference, we need to parse the generated linear output of the structured tree to obtain the final tuples of structured event records.
For evaluation, we first normalize the event records by converting them to lowercase format to avoid innocuous errors. 
Then we follow the same F1 metrics of four event elements as in the previous study \cite{metrics1,metrics3,Text2Event}, including Trigger Identification (\textbf{TI}), Trigger Classification (\textbf{TC}), Argument Identification (\textbf{AI}) and Argument Identification (\textbf{AC}).

\begin{table}[!t]
  \centering
\fontsize{8.5}{9.5}\selectfont
\setlength{\tabcolsep}{2.3mm}
\caption{
Overall results on sentence-level datasets.}
\vspace{-3mm}
  \label{tab:main-1}%
\begin{tabular}{lccccl}
\hline
\multicolumn{1}{c}{{\textbf{}}}
& {\textbf{TI}} 
& {\textbf{TC}}
& {\textbf{AI}}
& {\textbf{AC}}
& {\bf \emph{Avg}}\\
         \hline
\rowcolor{lightblue} \multicolumn{6}{c}{$\bullet$ \textbf{\emph{ACE05-EN\textsuperscript{+}}}} \\
Pipeline (Bart) &
60.8 	&57.0 &33.3 		&20.2 	 	&42.8 
\\
E2E (Bart)  & 
63.5 		&59.3 		&35.5 		&23.0 &45.3 \tiny{+2.5}
\\ 
\hdashline

Pipeline (T5)   & 
	61.2 		&57.1  	&33.1 	&20.4 		&43.0 

\\ 
E2E (T5)   	&\textbf{65.0} 	&\textbf{61.1}  &	\textbf{35.3}  &	\textbf{23.2} 	&		\textbf{46.2} \tiny{+3.2} \\ 

\hline

\rowcolor{lightblue} \multicolumn{6}{c}{$\bullet$ \textbf{\emph{ACE05-ZH}}} \\
Pipeline (mBart) & 
42.5&		29.7&	18.7&	15.8& 
	26.7 \\ 

E2E (mBart)  & 
43.3&	33.6&		19.9&		16.7&
	28.4 \tiny{+1.7} \\ 

\hdashline

Pipeline (mT5)   & 
43.9&		30.8&   17.6&	14.7&	26.8 

\\ 
E2E (mT5)    & 
	\textbf{44.2}	&	\textbf{34.5}&	\textbf{20.1}&	\textbf{17.3}&
\textbf{29.0} \tiny{+2.2}
\\

\hline

\rowcolor{lightblue} \multicolumn{6}{c}{$\bullet$ \textbf{\emph{PHEE}}} \\

Pipeline (Bart)  & 
	50.1& 	 	49.1 &		25.9& 	23.8 &		37.2 

\\ 
E2E (Bart)  &
\textbf{53.1} 		&\textbf{50.5}	&29.2 	&27.1 		&40.0 \tiny{+2.8}

\\

\hdashline

Pipeline (T5)   & 
	49.4 &		47.0 &	27.9 &		25.0 &		37.3 

\\ 
E2E (T5)    & 
 52.6 	&49.7  	&\textbf{32.2} 	&\textbf{29.9} 	&\textbf{41.1} \tiny{+3.8}

\\ 

\hline

\rowcolor{lightblue} \multicolumn{6}{c}{$\bullet$ \textbf{\emph{GENIA}}} \\
Pipeline (Bart) & 
23.5&		20.9&	-&	-& 	22.2

\\ 

E2E (Bart)  & 
	27.1&	24.3&		-&	-&
	25.7 \tiny{+3.5}
\\ 
\hdashline
Pipeline (T5)   &
	21.1&	18.3&		-&	-&
	19.7 

\\ 
E2E (T5)    & 
	\textbf{28.1}&		\textbf{25.3}&		-&	-&
	\textbf{26.7} \tiny{+7.0}

\\ 

\hline

\rowcolor{lightblue} \multicolumn{6}{c}{$\bullet$ \textbf{\emph{CASIE}}} \\
Pipeline (Bart) 
	&55.2  &	54.5 	&36.6 &32.9 		&44.8

\\ 

E2E (Bart) 
	&56.2  	&55.3 &\textbf{38.6} &\textbf{35.1} 	&\textbf{46.3} \tiny{+1.5}

\\ 
\hdashline
Pipeline (T5)   & 
	53.2 	&52.5 	&36.0	&31.9 	&43.4

\\ 
E2E (T5)    &
\textbf{56.5}	 	&\textbf{56.0}		&37.9 	&34.0 	&46.1 \tiny{+2.7}
\\ 
\hline
    \end{tabular}
    \vspace{-2mm}
\end{table}%

\begin{table}[!t]
  \centering
\fontsize{8.5}{9.5}\selectfont

   \setlength{\tabcolsep}{6pt} 
\caption{
Overall results on document-level datasets. }
\vspace{-3mm}
  \label{tab:main-2}%
\begin{tabular}{lccccl}
\hline
\multicolumn{1}{c}{{\textbf{}}}
& {\textbf{TI}} 
& {\textbf{TC}}
& {\textbf{AI}}
& {\textbf{AC}}
& {\bf \emph{Avg}}\\

         \hline
\rowcolor{lightblue} \multicolumn{6}{c}{$\bullet$ \textbf{\emph{RAMS}}} \\
Pipeline (Bart) & 
76.2& 	32.6& 	18.8& 	17.3& 	36.2 
\\ 

E2E (Bart)  & 

 \textbf{77.4} &  	36.7& 	20.9& 	19.3& 	38.6 \tiny{+2.4}

\\ 
\hdashline
Pipeline (T5)   & 
76.5& 	33.7& 	20.0& 	18.2& 	37.1

\\ 
E2E (T5)    &
76.8& 		\textbf{37.2}& 	\textbf{21.8}& 		\textbf{20.6}& 	
\textbf{39.1} \tiny{+2.0} 
\\ 

\hline
        
\rowcolor{lightblue} \multicolumn{6}{c}{$\bullet$ \textbf{\emph{WikiEvents}}} \\
Pipeline (Bart) & 
32.1&	29.0&	14.0&		10.8& 21.5 \\ 

E2E (Bart)  & 
35.3&		33.6&		17.4&		14.2&
25.1 \tiny{+3.6}

\\ 
\hdashline
Pipeline (T5)   & 
	32.3&	28.8&	12.8&		9.9&  21.0 

\\ 
E2E (T5)    & 
\textbf{35.8}&	\textbf{34.0}&		\textbf{18.0}&	\textbf{16.0}&
\textbf{26.0} \tiny{+5.0}
\\ 
\hline
\end{tabular}
\vspace{-4mm}
\end{table}%

\vspace{-1mm}
\subsection{Main Results of SpeechEE}
\label{Main Results of SpeechEE}

We present the main comparisons on different datasets under three scenarios, in Table \ref{tab:main-1}, \ref{tab:main-2} and \ref{tab:main-3}, respectively.
We note that these are testing results, where the models are trained on the mixture of the human-reading and synthesis training sets. 
According to the results, three key general observations can be found.
First, we see that our E2E SpeechEE method consistently achieves overall better results than the pipeline baseline among all different-level datasets. 
Besides, there is an effect of different pre-trained language models on the performance of SpeechEE, where T5 has stronger effects than Bart in most cases.
Lastly, for the four elements of EE, the recognition and classification of argument tend to be significantly lower than those of trigger, demonstrating the greater challenges faced by the Event Argument Extraction (EAE) task.
Following we summarize the detailed scenario-specific observations.

\paratitle{Sentence-level Performance.}
On ACE2005 dataset, both the pipeline and E2E methods exhibit a significant performance disparity between the Chinese and English datasets.
This is possibly due to that,
1) ASR model indeed performs better on English data than other languages; 
or 2) the performance of the language model differs in language as well.

\vspace{-2mm}
\paratitle{Document-level Performance.}

From the Table \ref{tab:main-2}, it can be seen that the RAMS dataset receives better results on the TI task, and the performance of the TC task decreases significantly, which is mainly because RAMS includes many more event types (139 types), which leads to poor performance on the TC task.
The WikiEvents data contains longer documents (speech over 5 minutes) and more events\&arguments with more complicated event schema, compared to RAMS dataset, where there thus are greater challenges of EE.

\paratitle{Dialogue-level Performance.}
The general pattern in Table \ref{tab:main-3} is mostly coincident with above results.
Overall, the performance on document and dialogue levels of SpeechEE is comparatively lower, leaving more room for improvement.

\begin{table}[t]
  \centering
\fontsize{8.5}{9.5}\selectfont

\setlength{\tabcolsep}{5pt} 
\caption{
Results on dialogue-level dataset. 
Duconv dataset has only one event type, thus no TC evaluation.
}

  \label{tab:main-3}%
\begin{tabular}{lccccl}
\hline
\multicolumn{1}{c}{{\textbf{}}}
& {\textbf{TI}} 
& {\textbf{TC}}
& {\textbf{AI}}
& {\textbf{AC}}
& {\bf \emph{Avg}} \\

         \hline

\rowcolor{lightblue} \multicolumn{6}{c}{$\bullet$ \textbf{\emph{Duconv}}} \\
Pipeline (mBart) & 
42.3&	-&	22.4&	20.2& 28.3

\\ 

E2E (mBart)  & 
45.4&	-&	23.8&	20.7& 30.0 \tiny{+1.7} 
\\ 
\hdashline
Pipeline (mT5)   &
43.1&		-&	22.2&	19.8& 28.4 

\\ 
E2E (mT5)    & 
\textbf{45.9}&		-&		\textbf{24.0}&		\textbf{21.1}&	\textbf{30.3} \tiny{+1.9}

\\ 

\hline

    \end{tabular}
    
\end{table}

\begin{table}[t]
  \centering
\fontsize{8.5}{9.5}\selectfont
\setlength{\tabcolsep}{2.7mm}

\caption{
Ablation results (F1) on PHEE dataset. 
SU: Shrinking Unit, 
CL: Contrastive Learning,
ED: Entity Dictionary.
}
\vspace{-2mm}
\begin{tabular}{lccccl}
\hline
\multicolumn{1}{c}{{\textbf{}}}
& {\textbf{TI}} 
& {\textbf{TC}}
& {\textbf{AI}}
& {\textbf{AC}} & {\bf \emph{Avg}}\\
\hline

E2E (T5) 	& 52.6	&49.7	&32.2&	29.9 & 41.1 

\\ 

\hdashline

\quad w/o SU&	52.1&	49.3	&31.5&	29.0 & 40.5 \tiny{-0.6} 

\\ 

\quad w/o CL&	51.0&	48.4&	31.0&	28.3 & 39.7 \tiny{-1.4}

\\ 

\quad w/o ED	&51.7	&48.7	&29.8	&27.6 & 39.5 \tiny{-1.6}

\\ 

\hline

    \end{tabular}
  \label{tab:ablation}%
  
\vspace{-1mm}
\end{table}%

\begin{figure}[!t]
    \centering
    \includegraphics[width=0.95\columnwidth]{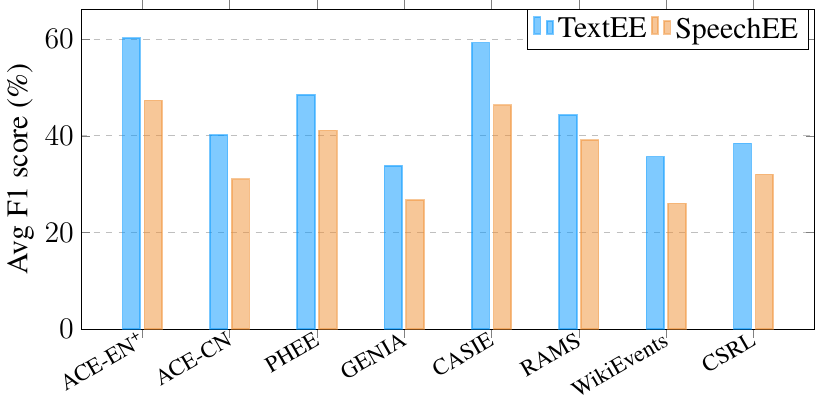}
    \vspace{-2mm}
    \caption{
    Comparisons between TextEE and SpeechEE.
    }
    \label{fig:TextEE}
    \vspace{-3mm}
\end{figure}

\vspace{-2mm}
\subsection{Model Ablation Study}

Here we provide ablation results on our advanced E2E system, to ground the exact contribution of each component and design within, i.e., Contrastive Learning (CL), Shrinking Unit (SU), and Entity Dictionary (ED).
We representatively select the sentence-level PHEE dataset with T5-large backbone.
As shown in Table \ref{tab:ablation}, we see that the performance of all four event elements drops persistently when any of the three modules is removed, which shows each of them is indispensable.
Specifically, the CL module gains better performance in the event detection task.
This indicates that the representation learning on speech and event features is of the most importance.
In contrast, the ED module plays a more role in the EAE task.

\begin{figure}[!t]
    \centering
    \includegraphics[width=0.99\columnwidth]{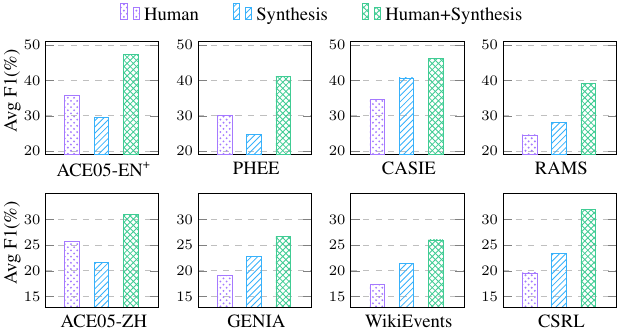}
    \vspace{-2mm}
    \caption{
    Analysis of the effect of synthesis dataset.
    }
    \label{fig:agument}
    \vspace{-3mm}
\end{figure}

\vspace{-2mm}
\subsection{Analysis and Discussion}

In this part, we delve deeper into our data and model, aiming to provide a more thorough understanding of them.

\paratitle{Q1: Is SpeechEE Really More Challenging Than TextEE?}
As a primary question, we aim to determine whether SpeechEE is meaningful and whether it presents greater challenges for the Event Extraction (EE) task compared to traditional textual EE. 
To this end, we compared the performance differences between SpeechEE and the corresponding TextEE. 
Using the same language model, T5-large, and adopting the identical generation-based E2E TextEE method across all eight datasets, the results are shown in Fig. \ref{fig:TextEE}. 
With a similar model (except for the encoder for the input signal), the performance of SpeechEE is generally lower than that of TextEE. 
This highlights the real challenges faced by the SpeechEE task.

\paratitle{Q2: Is It Necessary to Develop Synthesis Data?}
Next, we explore whether maintaining a synthesis dataset to enlarge the training corpus is meaningful to SpeechEE. 
Using the E2E model, we carry out experiments on all SpeechEE subsets under 3 varied training data: 1) only real data, 2) only synthesis data, and 3) mixture of two data.
The overall EE results (average F1) are shown in Fig. \ref{fig:agument}.
As seen, when comparing the performance using real data and synthesis data, the former setting performs largely better than the latter, such as on ACE05-EN\textsuperscript{+}, ACE05-ZH, and PHEE datasets. 
While for the rest datasets, the synthesis data yields better results. 
This is because the corresponding human-reading data are much smaller than synthesis data, where the former might not provide rich enough features for learning the pattern. 
Unsurprisingly, the system trained merely on human-reading data has a large decrease compared to the mixed training dataset.
This directly indicates that the synthesis data is effective in relieving the data scarcity issue for SpeechEE, even though the quality of the synthesis data might be inferior to the real speech.

\begin{figure}[!t]
    \centering
    \vspace{2mm}
    \includegraphics[width=0.95\columnwidth]{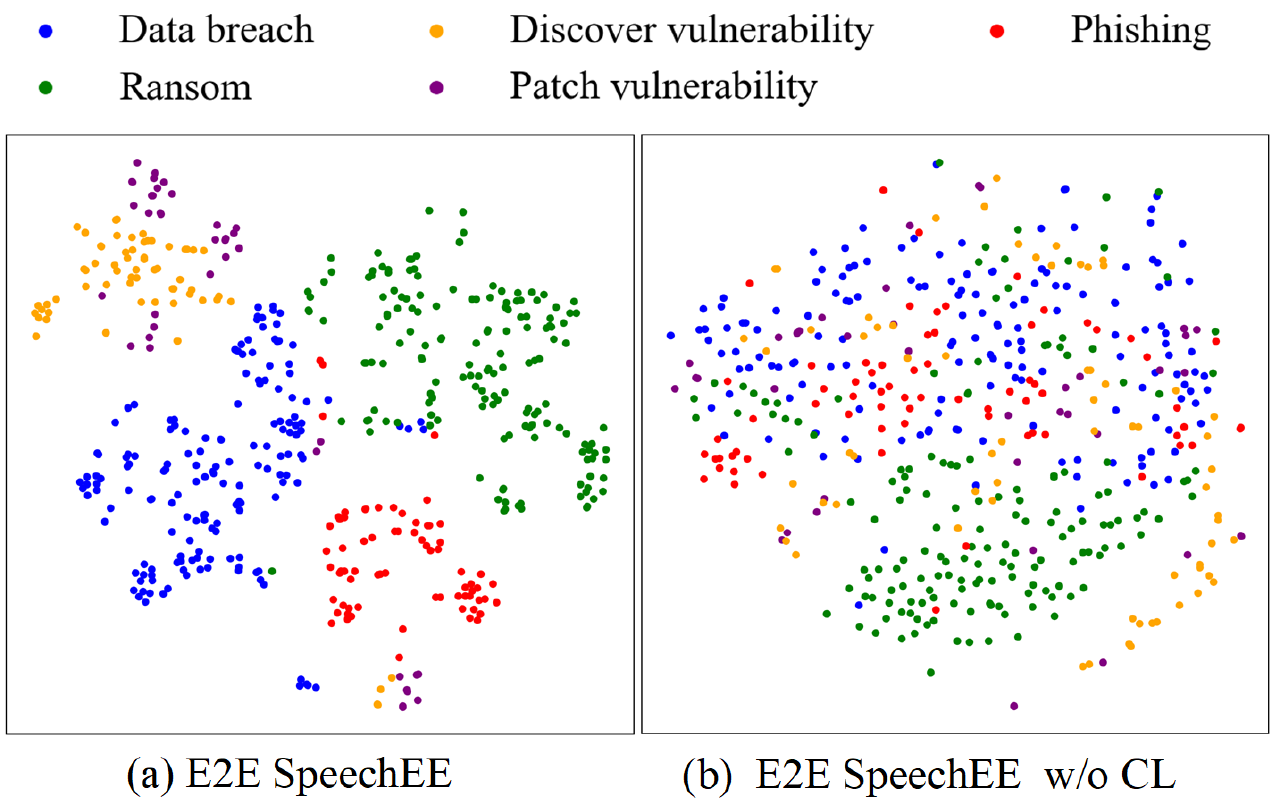}
    \caption{
    T-SNE visualization about the effect of CL. 
    }
    \label{fig:tsne}
    \vspace{-5mm}
\end{figure}

\begin{figure}[!t]
    \centering
    \includegraphics[width=0.9\columnwidth]{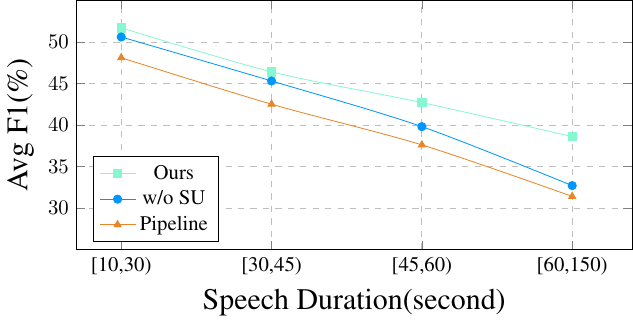}
    \vspace{-2mm}
    \caption{
    Impact of different speech durations on SpeechEE performance.
    }
    \label{fig:SU-analyze}
    \vspace{-4mm}
\end{figure}

\paratitle{Q3: How Does the Contrastive Learning Contribute?}
The above ablation study has demonstrated the efficacy of the Contrastive Learning mechanism in our system, in boosting the speech and event representations.
To reveal how it exactly improves the performance, here we present the visualization of the embeddings.
Technically, we randomly select 500 samples from the CASIE dataset labeled with 5 event types, and then obtain the speech encoder output embedding of them, and finally visualize the representations via t-SNE algorithm \cite{t-sne}.
From Fig. \ref{fig:tsne}, it is obvious that the samples of different event categories after Contrastive Learning have clearer boundaries than those without the mechanism, indicating that Contrastive Learning helps to learn event features from speech. 
In addition, we observe that among the 5 event types in the CASIE dataset, Patch-vulnerability and Discover-vulnerability show relatively poorer performance, which may be caused by the high semantic similarity between these two event types.

\begin{figure}           
    \centering
    \includegraphics[width=0.98\columnwidth]{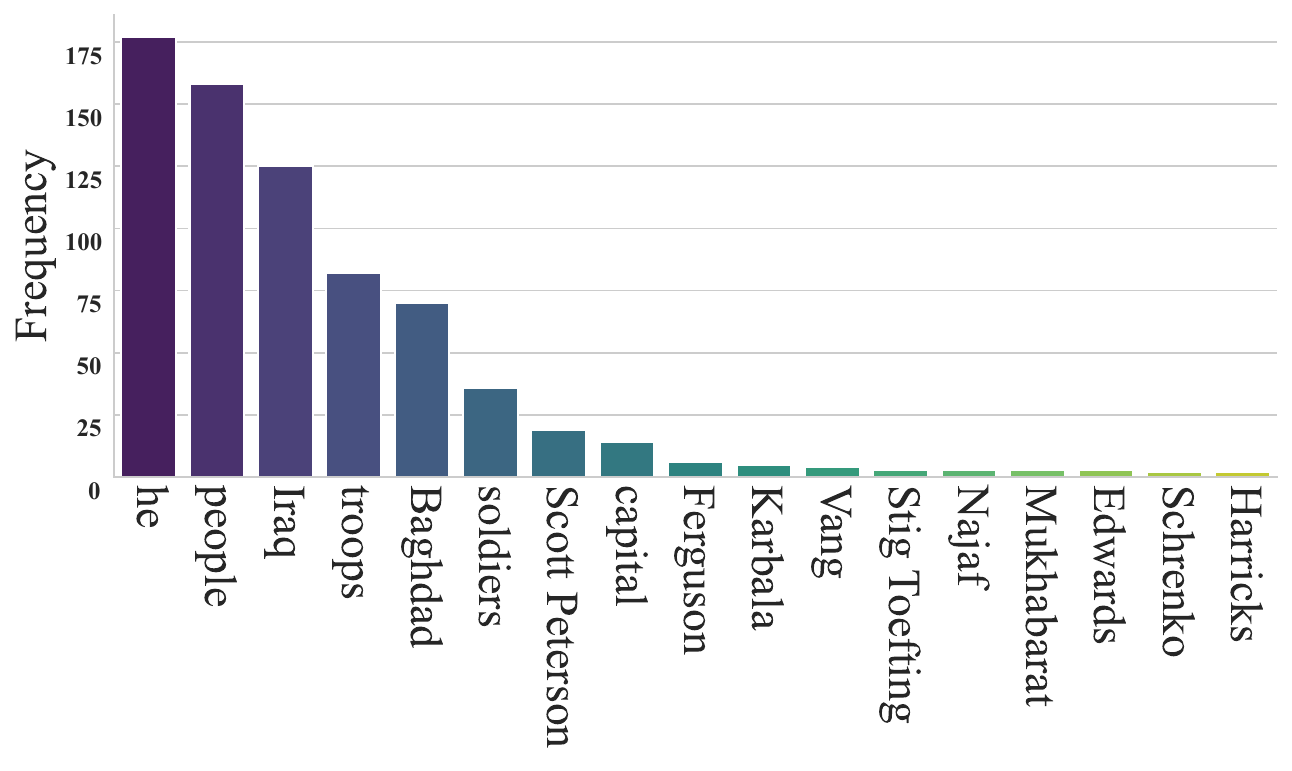}
    \vspace{-3mm}
    \caption{
    Frequent entities in ACE05-EN\textsuperscript{+} dataset. 
    }
    \vspace{-4mm}
    \label{fig:frequency}
\end{figure}

\begin{figure}[!t]
    \centering
    \includegraphics[width=0.53\columnwidth]{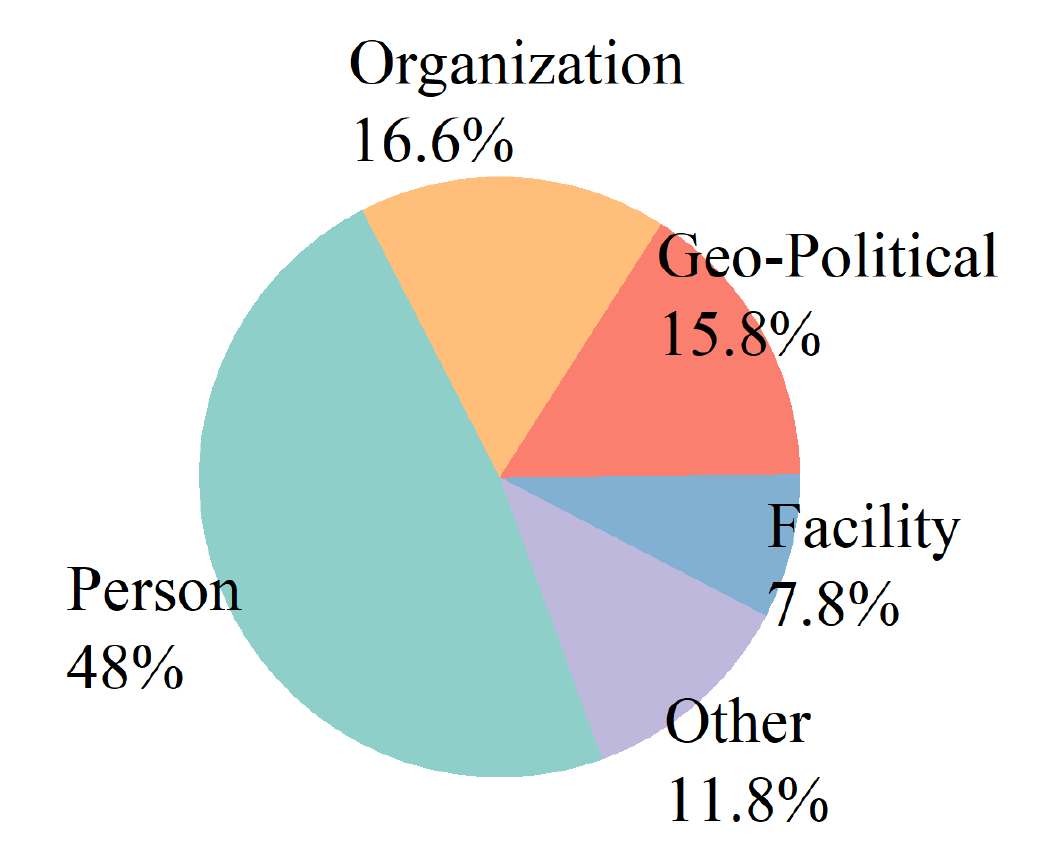}
     \vspace{-2mm}
    \caption{
   The component of Entity Dictionary.
    }
    \label{fig:pie}
    \vspace{-5mm}
\end{figure}

\paratitle{Q4: How Does the Shrinking Unit Module Address Lengthy Audio Signals?}
Speech signals are often much longer than text (especially in the form of long documents), which undoubtedly increases the modeling complexity of SpeechEE. 
We now evaluate our model's performance across different speech lengths. 
We consider the RAMS document data, where we grouped speech into four-length segments to observe the model's results. 
As illustrated in Fig. \ref{fig:SU-analyze}, as the length of the speech sequence increases, there is a significant decrease in the performance of all three systems. 
However, our 
E2E model, equipped with the full Shrinking Unit (SU) mechanism, effectively counters this trend, demonstrating its effectiveness. 
In contrast, our model without the SU experiences the most severe performance drop when the speech length exceeds 60 seconds, highlighting the crucial role of the module in handling long-duration speech.

\paratitle{Q5: Does Entity Dictionary Really Alleviate the Problem of Difficult Entity Extraction?}
To further explore how the proposed Entity Dictionary helps facilitate the task, we carry out an analysis study on the ACE05-EN\textsuperscript{+} dataset. 
We first count the frequency distribution of all the occurrences of entities in the train set, as shown in Fig. \ref{fig:frequency}, and we find that the distribution of entities is characterized by an obvious long-tailed distribution. 
Such data distribution characteristic makes the model tend to ignore the knowledge of uncommon entities during training and the model has difficulty generating words it rarely sees.
However, the Entity Dictionary helps focus on low-frequency entity information in the data source by introducing external knowledge. 
The component of the Entity Dictionary is mainly names of people, organizations, and places as shown in Fig. \ref{fig:pie}, which alleviates mistakes such as homophones, incorrectly extracting people's name ``\emph{Emmalie}'' as ``\emph{Emily}'' where the former is rarely seen while the latter is more common.

\begin{figure}[!t]
    \centering
    \includegraphics[width=0.99\columnwidth]{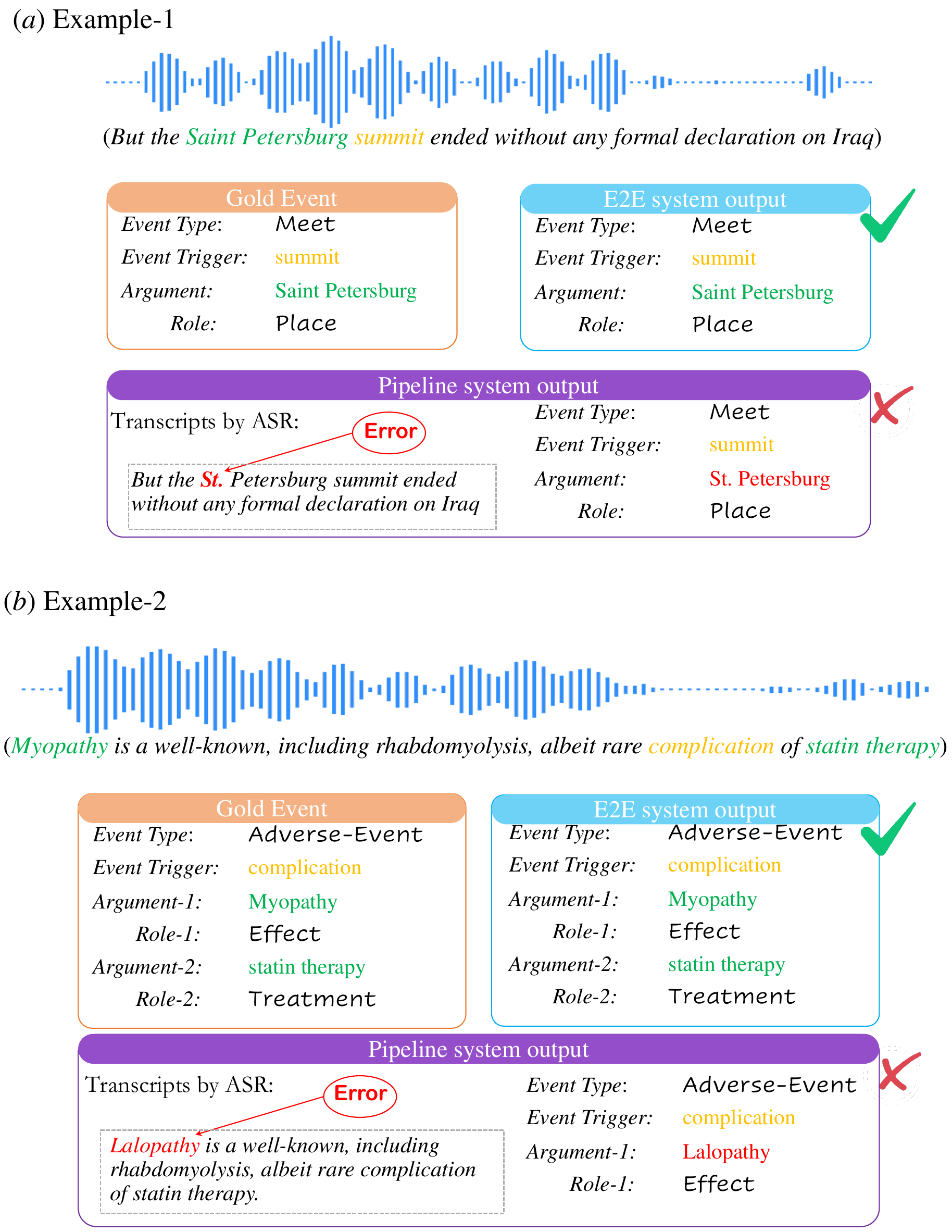}
    \vspace{-2mm}
    \caption{
    Qualitative examples of pipeline method and E2E method.
    }
    \label{fig:case study}
    \vspace{-5mm}      
\end{figure}

\vspace{-2mm}
\subsection{Qualitative Case Study}

Finally, we provide a more intuitive understanding of the differences in performance between our pipeline and E2E systems on specific instances by offering some qualitative case studies. 
We randomly select two samples from the sentence-level test set, where our E2E model correctly produced outputs that matched the gold events for both instances. 
However, the pipeline model fails in both cases, demonstrating typical errors.
For example-1, the pipeline system incorrectly recognized ``\emph{Saint Petersburg}'' as ``\emph{St. Petersburg}'' during the ASR stage (due to biases in the training of the ASR model). 
This error propagates through the system, leading to incorrect identification of the argument in the subsequent EE step.
For example-2, similarly, the ASR mistakenly identifies the word ``\emph{Myopathy}'' as ``\emph{Lalopathy}'', which results in incorrect event argument outcomes. 
Additionally, constrained by the two-step prediction paradigm, the pipeline system only identifies one argument, failing to recognize the second argument.

\section{What To Do Next with SpeechEE?}

We believe firmly that the proposed SpeechEE will open a new era for the multimodal IE community. 
Here we shed light on the potential directions for future research.

\paratitle{$\bullet$ Mitigating Noise Impact.}
Speech in real scenarios always includes background noise and other types of interference. Our experimental results also indicate that noisy backgrounds impose additional challenges on SpeechEE. We believe it is promising to develop stronger mechanisms to help filter out ambient noise in speech, enhancing task performance.

\paratitle{\(\bullet\) Identifying Implicit Elements.}
Beyond noise issues, SpeechEE often encounters implicit elements. While most EE results can find corresponding audio segments in speech, sometimes words are swallowed or not explicitly pronounced (termed as implicit elements). Compared to explicit elements, identifying implicit ones poses a greater challenge. We consider it crucial to devise smarter methods to address the recognition of these implicit elements.

\paratitle{$\bullet$ Cross-language SpeechEE.}
Our dataset includes two major languages, English and Chinese, with annotations that are not parallel across languages. Future research can explore cross-lingual transfer learning in speech, investigating the role of language-invariant features in enhancing EE task.

\paratitle{\(\bullet\) Weak/Unsupervised SpeechEE.}
In this paper, we primarily focus on supervised learning using a substantial amount of annotated data. We deem it essential to leverage our benchmark for weak or unsupervised SpeechEE. Some current multimodal large language models (MLLMs) \cite{zhang2024vpgtrans,wu24next,fei2024vitron,wu2024towards,fei2024enhancing,zhang2024omg} already exhibit significant unsupervised generalization capabilities. 
Future research can explore weak or unsupervised approaches in SpeechEE.

\paratitle{$\bullet$ Better Evaluation Metric for SpeechEE.}
The current evaluation method for the EE task strictly matches predicted event records with the golden label. However, given that the input for the SpeechEE task is purely audio without textual information, strict matching significantly hinders performance. A new evaluation metric that accommodates fuzzy semantic matching is expected to be proposed for a fair evaluation of SpeechEE. For example, an entity that semantically matches the core meaning with the gold label should be considered correct.

\section{Conclusion}
\label{sec:Conclusion}

In this paper, we introduce a novel task, SpeechEE, extracting structured event information from speech. 
We first contribute a comprehensive dataset tailored to this task, which features diverse scenarios, languages, domains, and speaker styles, constructed from both synthesis and human reading. 
Further, we propose an E2E SpeechEE model to offer a strong baseline for the task. 
Through analysis, we demonstrate the complexity of the task, and the effectiveness of our approach. 
Finally, as pioneers in this topic, we highlight key directions for future research.

\begin{acks}
This work is supported by the National Natural Science Foundation of China (NSFC) Grant (No. 62336008).    
\end{acks}

\bibliographystyle{ACM-Reference-Format}
\balance
\bibliography{acmmm}

\clearpage

\appendix

\section{Potential Ethical Considerations}

We conduct all procedures for data collection and annotation by ethical principles and with the informed consent of the participants.

\paratitle{\(\bullet\) Privacy Concerns.}
In developing the SpeechEE dataset, we prioritized privacy and ethical data use by meticulously ensuring that the combined datasets did not contain any personally identifiable or sensitive information. Our dataset, derived from existing textual EE datasets, represents a diverse array of news types, and we have made certain that all data used are legally sourced and devoid of any individual-specific details. 
Additionally, to further protect privacy, the dataset is publicly available, allowing for transparency and accountability in its use. 
These precautions are designed to prevent any potential misuse of personal information while fostering a secure environment for academic and applied research.

\paratitle{\(\bullet\) Annotator and Compensation.}
Recognizing the crucial role of human annotators in developing our dataset, we employed highly skilled senior postgraduate students trained specifically for these tasks. The time required to annotate each segment of the dataset is substantial, typically ranging from 3 to 5 minutes per segment due to the complexity and precision required. To fairly compensate for their effort and expertise, annotators are paid 1 yuan (approximately \$0.15 USD) for each segment they annotate. Moreover, the compensation scheme for linguistics and computer science experts who contribute to our project is carefully calibrated to reflect their time investment and expertise, ensuring equitable remuneration. This approach underlines our commitment to ethical practices in compensating all contributors fairly for their labor and intellectual input.

\paratitle{\(\bullet\) Consent and Transparency.}
In the creation of datasets, it is paramount that all contributors—whether their participation involves providing speech samples directly or indirectly—are fully informed about how their data will be used and have explicitly consented to it. Transparency about the data collection process and the intended use of the data is crucial to maintaining ethical standards.

\paratitle{\(\bullet\) Bias and Fairness.}
Given that the dataset comprises diverse scenarios, languages, and speaker styles, it is important to systematically analyze and address any potential biases that may be present. These biases could manifest in the form of underrepresentation of certain languages, dialects, or demographic groups. Efforts should be made to ensure the SpeechEE system does not perpetuate or amplify existing biases.

\paratitle{\(\bullet\) Impact on Employment.}
The automation of EE from speech could potentially impact jobs that traditionally rely on human transcription and analysis, such as secretarial or journalistic roles. It is important to consider the broader socio-economic implications of this technology and engage with affected stakeholders to explore supportive measures, such as retraining programs.

\paratitle{\(\bullet\) Misuse Potential.}
Lastly, the capability to automatically extract structured information from speech could be misused in scenarios like surveillance without consent, eavesdropping, or other forms of intrusion into personal or confidential communications. Strong guidelines and possibly regulatory frameworks should be proposed to prevent misuse of this technology, ensuring it is used ethically and responsibly.

\section{Extended Model Implementation and Settings}
\label{app:Model and Setup Specification}

Here, we provide additional technical details about the model as an expansion of the main paper.

\subsection{More Model Details on Pipeline SpeechEE Method}

\paratitle{ASR Model.}
The whisper model is chosen as our ASR module for its high capability in learning speech features and outstanding ASR performance. 
In comparison to alternative ASR tools such as Wav2Vec 2.0 \cite{w2v2} and HuBERT \cite{HuBERT}, the whisper model undergoes training on a substantially labeled audio corpus (approximately 680,000 hours), a volume approximately ten times larger than the pre-training data (60,000 hours) utilized for unsupervised tasks like masked prediction in Wav2Vec 2.0. Consequently, it is capable of directly acquiring the mapping from speech to text, thereby exhibiting superior performance in speech recognition compared to other ASR models. Moreover, the whisper model integrates data from diverse languages and domains, aligning with the multilingual and varied domain characteristics of the SpeechEE dataset we have constructed.

The whisper model consists of three parts: an acoustic feature extraction module, a transformer encoder, and a decoder.

\begin{itemize}
    \item 
    Feature Extraction: Given a speech, the acoustic feature extraction module aims to get a log Mel spectrogram representation with 80 channels by using a 25 ms window and a 10 ms step.

    \item 
    Encoder: After 2 one-dimensional convolutional layers and
GLUE activation function for length reduction, the 
audio representation is fed into 12 layers of transformer modules to encode the acoustic features and get the encoder output last hidden state.

    \item 
    Decoder: The decoder uses the same number of transformer modules as the encoder. The last hidden state of the encoder is fed into the decoder through a cross-attention mechanism. Then the decoder autoregressively predicts textual tokens based on the hidden state and previously predicted tokens.
    
\end{itemize}

\paratitle{TextEE Model.}
Text2Event is a generative-based E2E EE method where the entire EE process is modeled uniformly in a sequence-to-structure architecture. All trigger words, arguments, and their type labels are generated uniformly as natural language words to extract events from text in a direct manner.

\begin{itemize}
    \item 
    \textbf{Why do we choose Text2Event for the SpeechEE task?} Text2Event method is chosen for the SpeechEE task because it is data-efficient which means it can be learned using only coarse parallel text-record annotations, i.e. <sentence, event records>, instead of fine-grained token-level annotations. That matches the SpeechEE task perfectly where fine-grained trigger and argument mention annotation is absent because the speech signal is boundless. Therefore, following Text2Event, we adopt the generated-based EE method as the TextEE module of pipeline SpeechEE.

    \item 
    \textbf{Decoding Strategy.} Different from the greedy decoding strategy where the model tends to select the token with the highest predicted probability at each decoding step, a trie-based decoding strategy is used in the SpeechEE decoder module.
This is because a greedy decoding algorithm can not guarantee to generate valid event structures. 
In other words, it may lead to invalid event types, mismatched argument types, and incomplete structures. 

In addition, the greedy decoding algorithm ignores useful knowledge of event patterns that can effectively guide the decoding.
In the SpeechEE model, a trie-based constraint decoding method dynamically selects and prunes a candidate vocabulary based on the currently generated state. 
The candidate vocabulary consists of event schema, mention to extract, and structure indicator. 
The event schema, including the event type and the argument role bonding to the event type, is injected into the decoder as external event knowledge to realize the controllable event record generation.
``Event Type'' and ``Argument Role'' are pre-defined and serve as constrained candidate vocabulary at the certain decoding step to guarantee the correctness of the event scheme. 
\end{itemize}

\subsection{Model Implementation Configurations}

\paratitle{Evaluation Metrics.}
We evaluate the SpeechEE model by using four subtasks in EE.
\begin{itemize}
    \item Trigger Identification (\textbf{TI}) : A trigger is correctly identified if it matches a reference trigger. 
    \item 
Trigger Classification (\textbf{TC}) : A trigger is correctly classified if its event type and trigger mention match the reference.
    \item 
Argument Identification (\textbf{AI}) : An argument is correctly identified if its event type and argument mention match a reference argument. 
    \item 
    Argument Identification (\textbf{AC}) : An argument is correctly classified if its event type, argument role and argument mention all match a reference argument. 
\end{itemize}

\paratitle{Model Configurations.}
We use the pre-trained whisper-large-v2 encoder, T5-large decoder, and Bart-large decoder for the E2E SpeechEE model.
For efficient training, we freeze the acoustic feature extraction module of whisper and train the self-attention encoder, Shrinking Unit module, cross-attention between encoder and decoder, and Entity Dictionary attention. 
We train all SpeechEE models for 30 epochs and optimize the models using AdamW \cite{Adamw}.
We conduct experiments on NVIDIA A100 80GB. 
All our models are evaluated on the best-performing checkpoint on the validation set.
The detailed hyperparameters setting is shown in Table \ref{tab:hyperparameter}.

\section{Extended Data Specification}
\label{app:Extended Data Specification}

\subsection{Data Source Description}

The raw data is from well-known textual EE datasets, including five sentence-level datasets ACE05-EN\textsuperscript{+} \cite{metrics3}, ACE05-ZH \cite{ACErawdata}, PHEE \cite{PHEE}, CASIE \cite{CASIE} and GENIA \cite{GENIA}; two document-level datasets RAMS \cite{RAMS} and WikiEvents \cite{Wiki}; one dialogue-level subset Duconv in CSRL \cite{CSRLdata}, which is revised from Semantic Role Labeling task similar to EE.

    \textbf{ACE05-EN\textsuperscript{+}}\footnote{\url{https://blender.cs.illinois.edu/software/oneie/}} is a benchmark dataset for event extraction in the English language. It is created as part of the Automatic Content Extraction (ACE) program. The genres include newswire, broadcast news, broadcast conversation, weblogs, discussion forums, and conversational telephone speech. 
    Particularly, for the ACE05-EN\textsuperscript{+} dataset, we follow the same split and preprocessing step with the previous work \cite{metrics3}, which extended the original ACE05-EN data by considering multi-token event triggers and pronoun roles and marked it with a `+'.

    \textbf{ACE05-ZH}\footnote{\url{https://catalog.ldc.upenn.edu/LDC2006T06}} is a Chinese version of the ACE05 dataset. It is developed to facilitate research in event extraction from Chinese texts. Like ACE05-EN, it contains annotated data for training and testing, covering 33 event types. ACE05-ZH plays a crucial role in advancing event extraction research for the Chinese language domain.

     \textbf{PHEE}\footnote{\url{https://github.com/zhaoyuesun/phee}} is a dataset for pharmacovigilance comprising over 5000 annotated events from medical case reports and biomedical literature. It is designed for biomedical event extraction tasks.

     \textbf{CASIE}\footnote{\url{https://github.com/Ebiquity/CASIE}} is an event extraction dataset focusing on the cybersecurity domain. The corpus contains 1000 annotations and source files. It annotates event instances with rich annotation and defines five event types in the cybersecurity domain: Databreach, Phishing, Ransom, Discover, and Patch.

     \textbf{GENIA}\footnote{\url{https://github.com/openbiocorpora/genia-event}} is the primary collection of biomedical literature compiled and annotated within the scope of the GENIA project. It contains 1,999 Medline abstracts, selected using a PubMed query for the three MeSH terms “human”, “blood cells”, and “transcription factors”. The corpus has been annotated with various levels of linguistic and semantic information. We follow the data processing approach as \cite{InstructUIE} and the GENIA dataset is only used for event detection tasks with five event types.

       \textbf{RAMS}\footnote{\url{https://nlp.jhu.edu/rams/}} is a widely used document-level event extraction dataset. It contains 9,124 annotated events from news based on an ontology of 139 event types and 65 roles. In the RAMS dataset, the document is relatively short, and each document is annotated with one event.

    \textbf{WikiEvents}\footnote{\url{https://github.com/raspberryice/gen-arg}} is a document-level event extraction benchmark dataset that includes complete event and coreference annotation. The corpus is collected from English Wikipedia articles that describe real-world events and then follow the
reference links to crawl related news articles. It contains 3,241 events in total, covering 50 event types and 59 argument roles. 

      \textbf{Duconv}\footnote{\url{https://github.com/syxu828/CSRL_dataset}} is a dialogue-level dataset which originates from the conversational semantic role labeling task. It contains 3,000 dialogue sessions focusing on movies and stars domain.
    We format this dataset for the EE task where the predicate works as the trigger in the EE task and the arguments of the predicate work as the argument mentions. Due to the predicate having no fine-grained classification, we consider Duconv as an EE dataset with only one event type and eight argument roles.

\begin{table}[t]
  \centering
  \caption{
The main hyperparameters of the model.
}
\fontsize{9}{11}\selectfont
\setlength{\tabcolsep}{5mm}
\begin{tabular}{lc}
\hline

 {\textbf{Hyperparameters}} &  \bf Value \\
\hline


epoch	&30 \\
learning rate &	5e-5 \\ 
batch size	& 16 \\
convolutional layer &	2\\
kernel size	&3 \\
stride &	2\\

\hline

    \end{tabular}

  \label{tab:hyperparameter}%
\end{table}%

\subsection{Specification on Data Construction}
\label{app:Specification on Data Construction}

Based on the textual EE dataset, we employ a method of manual reading to record corresponding speech signals. Considering the high cost associated with manual recording, we conduct random sampling from each textual EE dataset according to the original scale. 
Specifically, we randomly sample 1,000 instances for ACE05-EN\textsuperscript{+}, ACE05-ZH, GENIA, and RAMS datasets, 500 instances for PHEE and CASIE datasets, 140 dialogues for the Duconv dataset and 100 long documents for WikiEvents dataset.

\paratitle{Human-reading Speech Construction.}
To ensure the diversity of languages and speaker styles in the human-reading SpeechEE dataset, we have separately enlisted the assistance of 10 native speakers for both the Chinese and English languages. 
These speakers encompass a wide range of ages, genders, and tones, contributing to the diverse styles present in the human-reading SpeechEE dataset.
During recording sessions, participants are instructed to maintain a distance of 25 cm between their phones and their mouths, ensuring clarity and accuracy in reading the EE text within a quiet room.
Furthermore, to emulate real-life scenarios, we have also recorded the speech that includes background noise. In these instances, speakers are prompted to record their speech amidst various environmental noise.
The noisy background covers ten different scenarios, including street, airport, classroom, cafeteria, supermarket, stadium, meeting room, office, pet store, and rainy days outdoors. A detailed description of the environment settings is shown in Table \ref{tab:ambience}.
Following the recording process, two experts meticulously conduct cross-inspections to uphold the high quality of the human-reading speech data.

\begin{table*}[t]
  \centering
  \caption{
Environment settings on the speech ambiances.
}
\fontsize{9}{11}\selectfont
\setlength{\tabcolsep}{5mm}
\begin{tabular}{ll}
\hline

 \textbf{Environment} &  \textbf{Description} \\
\hline

Street	& Busy city streets with pedestrians passing and car noise.\\
Airport & Crowded airports with broadcasting and crowd noise.\\

Classroom &	Classrooms where students have lively discussions during the break of the class. \\ 
Cafeteria	& The cafeteria with the clattering of dishes and utensils and the chatting noise of people. \\
Supermarket & Busy supermarkets with the shoppers talking noise and the sound of cash registers.\\
Stadium &  School stadiums with the sound of sports and cheering.\\
Meeting room &	Meeting rooms with multi-speakers talking and arguing.\\
Office  &  Working offices with the noise of printers, photocopiers, ringing telephones, and conversations of workers.\\
Pet store &  Pet stores with the bark of animals and chatting noise of customers.\\
Rainy days outdoors	& Rainy days with the noise of wind and rain outdoors.\\

\hline

    \end{tabular}

  \label{tab:ambience}%
\end{table*}%

\paratitle{Automatic Speech Synthesis.}
First, we introduce the two TTS systems and explain the rationale behind our choice. 
Bark, a transformer-based TTS model developed by SunoAI, can generate highly realistic, multilingual speech, along with additional audio features such as music, background noise, and simple sound effects. 
Furthermore, the model is capable of simulating nonverbal expressions like laughter, sighing, and crying. 
Given its exceptional performance in generating English speech, we prefer Bark as the TTS model for sentence-level English SpeechEE texts because Bark can't support synthesizing long texts into speech.
Therefore, for longer English passages and Chinese text, we employ Edge-TTS as the TTS tool, mainly because of its ability to produce lengthy content and its relatively high quality in generating Chinese speech.

When building the automatic synthesis speech, we also take the speaker's style diversity into full account. So, we use different voice presets in the TTS system to control the synthesized intonation. Specifically, for the Bark TTS system, we use 10 different English speaker voices; for the Edge-TTS system, we use 7 different English speaker voices and 6 Chinese speaker voices. Both TTS systems cover male and female voices to guarantee the diversity of genders and tones.
In addition, we also manually add some background noise into the raw synthesis speech. The ambiences include the 
cafeteria, meeting room, street, classroom, and rainy days outdoors, which cover many real-world speech scenarios.

After using the TTS model to convert the text of the textual EE train set to audio, we reconstruct the dataset as speech-event records pairs for the SpeechEE task and filter the nonsense instances that only contain meaningfulness information for EE such as ``\#\#20010615'', ``...um'' because these meaningfulness instances may generate abnormal audios such as empty cases, which will cause error for the following speech processing.
Finally, we construct the SpeechEE synthesis dataset illustrated in Table \ref{tab:statistics}.
The synthesis data will be used to help augment the training corpus for better modeling SpeechEE task.  

\begin{figure}[!t]
    \centering
    \includegraphics[scale=0.3]{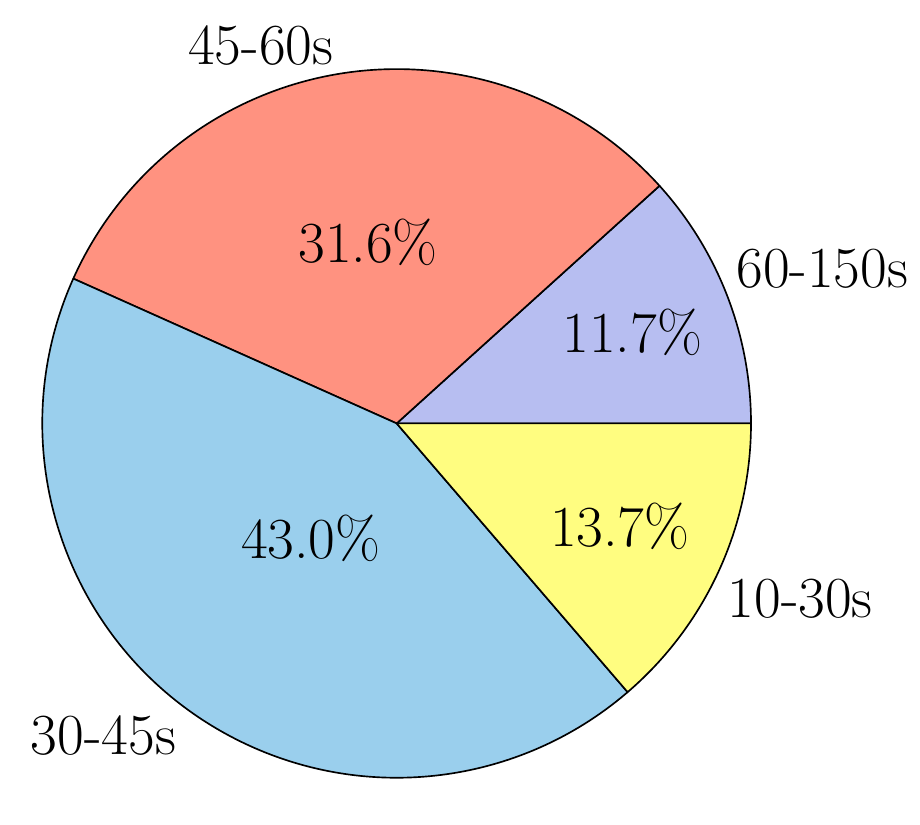}
    \caption{
    Speech duration statistics of RAMS dataset.
    }
    \label{fig:RAMS}
 
\end{figure}

\paratitle{Quality Control.}
As for quality controlling, we employ two indicators to show the data quality, the objective indicator, and the subjective indicator. 
For the objective indicator, we mainly focus on the accuracy of synthesis speech.
Therefore, a SOTA ASR model is used to evaluate the word error rate (WER) of the synthesis data.
The ASR model we choose is whisper-large-v2. 
The average WER for the English corpus and average CER for the Chinese corpus are $11.4\%$  and $7.2\%$ respectively, which shows the synthesis speech can keep most of the semantic information. 
For the subjective indicator, we consider naturalness as the quality indicator of synthesis speech.
So we launch a listening test to evaluate our SpeechEE dataset.
Two experts are recruited and asked to rate the naturalness of the synthesis speech using the 10-scale mean opinion score. 
The Cohesion Kappa score is utilized to assess the level of agreement among experts, measuring their consistency. Following the standard quality control process, we compute the Cohesion Kappa score, attaining 0.8 for the synthesis SpeechEE dataset, indicative of the good quality of our datasets. 

\subsection{Specification on SpeechEE Duration}
For sentence-level datasets, all synthesis datasets except ACE05-ZH are generated by the Bark TTS model, which is limited from 1 second to 15 seconds.
For the RAMS dataset in which the speech duration differs obviously from 10 seconds to three minutes, we count the speech duration distributions and show them in Fig. \ref{fig:RAMS}. 
RAMS dataset is mainly composed of speech within 60 seconds, which shows its documents are shorter but more numerous (9,124 documents).
In contrast, the WikiEvents dataset consists of 246 long documents which average 4 minutes duration. These two document-level datasets are quite complementary when evaluating the SpeechEE model under different cases.

\subsection{Specification on SpeechEE Data Insight}

\paratitle{$\bullet$ Two Construction Approaches:}
Our SpeechEE dataset contains two construction approaches: manually crafted human-reading speech, and synthesized speech generated using high-performance TTS systems. 
On one hand, human-reading speech includes rich information from real-life scenarios such as emphasis, pauses, onomatopoeia, and emotional cues. 
On the other hand, synthesized speech serves as an efficient data augmentation method, aiding in the rapid construction of large-scale training corpora and addressing the high cost associated with manually constructing datasets.

\paratitle{$\bullet$ Multiple Scenarios:} Our SpeechEE dataset covers three major common scenarios of existing EE: sentence-level, document-level, and dialogue-level, which can help to evaluate the performance comprehensively.

\paratitle{$\bullet$ Multiple Languages:} Our SpeechEE dataset covers two languages, Chinese for ACE05-ZH and 
Duconv datasets and English for the other six datasets.

\paratitle{$\bullet$ Diverse Domains:} Our SpeechEE dataset covers a wide range of topics and fields including news reports, medicine effects, genetic biology, cybersecurity, movies and stars, and other general domains. The diverse domains enable models to be trained and applied in a wider range of real-world scenarios.

\paratitle{$\bullet$ Rich Tones and Genders:} Our SpeechEE dataset also considers the diversity of speaker styles. Specifically, we have 17 English speaker voices and 6 Chinese speaker voices in the synthesis speech by choosing different TTS voice presets. For human-reading speech, we recruit 10 different native speakers of English and Chinese respectively. These voices include men and women and differ in speech volume, speed, and intonation.

\paratitle{$\bullet$ Different Ambiences:} Our SpeechEE dataset considers two background settings, including the quiet background and various noisy scenarios in the real world.
The ambiences include 10 different background settings such as car noise in the street, crowd noise in the cafeteria and classroom, multi-speaker noise in the meeting room, and rainy day noise outdoors, which can be found in Table \ref{tab:ambience}.

\paratitle{$\bullet$ Large Scale and High Quality:} Our SpeechEE dataset not only contains the manually crafted human-reading speech, but we also augment it with synthesis speech by using TTS systems to enlarge its scale by a factor of 10. At the same time, strict human cross-inspection is conducted to ensure the high quality of the whole speech data.

\end{document}